\documentclass{WileyChemistry-template}

\usepackage[T1]{fontenc}    %

\usepackage[usedoi=true,linkdoi=true]{rsc}

\usepackage[]{hyperref}

\usepackage{amsmath}     %

\newcommand{\eg}{e.\,g.}
\newcommand{\ie}{i.\,e.}

\newcommand{\rcm}{\ensuremath{\text{cm}^{-1}}}

\newcommand{\rateU}{\ensuremath{\text{cm}^{3}\,\text{s}^{-1}}}

\newcommand{\massU}{\ensuremath{m/z}}
\newcommand{\tK}{\ensuremath{\text{K}}}

\newcommand{\cmsmol}{\ensuremath{\text{cm}^2 \cdot \text{mol}^{-1}}}

\newcommand{\HCOp}{\ce{HCO+}}
\newcommand{\HOCp}{\ce{HOC+}}

\newcommand{\iNN}{\ce{^{15}N2}}

\newcommand{\HCOOp}{\ce{HCO2+}}

\newcommand{\COO}{\ce{CO2}}

\newcommand{\ArHp}{\ce{ArH+}}

\newcommand{\HH}{\ce{H2}}

\newcommand{\vv}{\ensuremath{2\,\nu_1}}
\newcommand{\vnu}{\ensuremath{2\,\nu_1}}
\newcommand{\vvone}{\ensuremath{1\,\nu_1}}
\newcommand{\vvnull}{\ensuremath{0\,\nu_1}}
\newcommand{\VNU}{\ensuremath{20^00 \leftarrow 00^00}}
\newcommand{\Vvv}{\ensuremath{20^00}}

\newcommand{\Wvv}{\ensuremath{\omega_{2\nu_1}}}
\newcommand{\Bvv}{\ensuremath{B_{2\nu_1}}}
\newcommand{\Dvv}{\ensuremath{D_{2\nu_1}}}
\newcommand{\Hvv}{\ensuremath{H_{2\nu_1}}}

\newcommand{\Tr}{\ensuremath{\tau_\mathrm{rad}}}
\newcommand{\Tq}{\ensuremath{\tau_\mathrm{q}}}
\newcommand{\kq}[1]{\ensuremath{k_{q\ce{#1}}}}

\newcommand{\TTT}[1]{\textsuperscript{{[#1]}}}
\newcommand{\MMM}[1]{\hyperref[#1]{Methods}}

\usepackage{natbib}
\setcitestyle{super, square, sort&compress, comma}

\title{\raggedright Overtone Transition $2\nu_1$ of HCO$^+$ and HOC$^+$: Origin, Radiative Lifetime, Collisional Quenching}

\author{
\begin{minipage}{\textwidth}
	Miguel Jim{\'e}nez-Redondo,\textsuperscript{[a]}\textsuperscript{+} Liliia Uvarova,\textsuperscript{[b]}\textsuperscript{+} 
  Petr Dohnal,\textsuperscript{[b]} Miroslava Kassayov{\' a},\textsuperscript{[b]}
	Paola Caselli,\textsuperscript{[a]} Pavol Jusko$^*$\textsuperscript{[a]} 
\end{minipage}
}

\newcommand{\affiliation}{
\begin{itemize}   
\item[{[a]}] Dr. Miguel Jim{\'e}nez-Redondo, prof. Dr. Paola Caselli, Dr. Pavol Jusko$^*$\\
Max Planck Institute for Extraterrestrial Physics, Gießenbachstraße 1, 85748 Garching, Germany\\
E-mail: pjusko@mpe.mpg.de\\

\item[{[b]}] Liliia Uvarova, Dr. Petr Dohnal, Miroslava Kassayov{\' a}\\
Department of Surface and Plasma Science, Faculty of Mathematics and Physics, Charles University, V Hole{\v s}ovi{\v c}k{\' a}ch 2, Prague 18000, Czech Republic\\
\end{itemize}
}

\renewcommand{\abstract}{
We present spectra of the first overtone vibration transition of C--H/ O--H stretch (\vv) in \HCOp\ and \HOCp,
recorded using a laser induced reaction action scheme inside a cryogenic 22 pole radio frequency trap.
Band origins have been located at 6078.68411(19) and 6360.17630(26)~\rcm, respectively.
We introduce a technique based on mass selective ejection from the ion trap for recording 
background free action spectra. Varying the number density of the neutral action scheme reactant (\ce{CO2} and Ar, respectively) 
and collisional partner reactant inside the ion trap, permitted us to estimate the radiative 
lifetime of the state to be 1.53(34) and 1.22(34) ms, respectively, 
and the collisional quenching rates of \HCOp(\vv) with He, \ce{H2}, and \ce{N2}.
}

\newcommand{\keywords}{
    astrophysics \textbullet\
	cryogenic ion trap \textbullet\ 
	ion-molecule interactions \textbullet\ isomers \textbullet\
    vibrational spectroscopy 
}

\begin{document}

\twocolumn[\vspace{-1.5cm}\maketitle\vspace{-1cm}
	\textit{\dedication}\vspace{0.4cm}]
\small{\begin{shaded}
		\noindent\abstract
	\end{shaded}
}

\begin{figure} [!b]
\begin{minipage}[t]{\columnwidth}{\rule{\columnwidth}{1pt}\footnotesize{\textsf{\affiliation}}}\end{minipage}
\end{figure}

\section*{Introduction}
\label{introduction}

The formyl cation, \HCOp, is an abundant molecule in the interstellar medium, 
being first detected already more than 50 years ago, 
interestingly under the name ``X-ogen''\cite{Buhl1970}, prior to its laboratory identification\cite{Woods1975}.
Although the existence of its higher energy isomer isoformyl cation, \HOCp, was clear, its
microwave spectrum was only measured in a dc glow discharge a decade later, \cite{Gudeman1982} 
and its interstellar detection quickly followed.
The study by \citet{Woods1983} focused on searching for \HOCp\ in 14 interstellar sources,
successfully detected \HOCp\ towards Sgr B2, and was even able to estimate the ratio of [\HCOp]/[\HOCp] $\approx$ 375.
\HCOp/ \HOCp, together with \ce{HCN}/ \ce{HNC}, are the two most simple linear closed shell isomeric systems, 
playing a crucial role in astrophysics, as
well as in fundamental molecular physics, \ie, properties of molecular ions. 
\begin{figure}
\centering
\includegraphics[width=80mm]{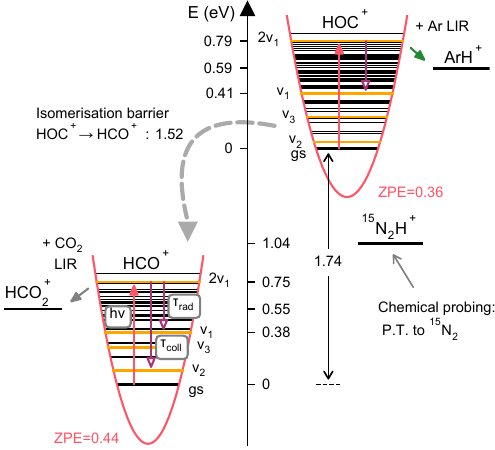}
\caption{Energy level diagram for \HCOp\ and \HOCp\ ions. 
    The vibrational energy levels are calculated from \HCOp\ \cite{Neese2013} and \HOCp\ \cite{Martin1993}
    spectroscopic constants, the zero point energies are taken from ref. \cite{Koput2019}.
    The red arrows represent the induced photon excitation, the magenta arrows radiative 
    and collisional deexcitation.  
    The gray, green arrows represent the LIR reaction pathways to \ce{HCO2+} and \ce{ArH+}, respectively.
    The \HCOp\ does not proton transfer to \iNN, unless sufficiently excited. See text for details.  
}
\label{fig_levels}
\end{figure}
Of the two isomers, \HCOp\ is energetically more favourable compared to \HOCp, with the ground states separated by 1.74 eV 
(from proton affinities on O and C, respectively\cite{Freeman1987b}), 
and the isomerisation barrier from $\HOCp \to \HCOp$ is 1.52 eV \cite{Chalk1997}.
The energy diagram of levels within both isomers can be seen in Fig.~\ref{fig_levels}. 

The \citet{Woods1975} study was the first to perform ground rotational spectroscopy on \HCOp\ in a discharge. 
IR spectroscopy, first determining $\nu_1$ \cite{Gudeman1983}, then also reporting rotationally resolved 
P, R branches therein \cite{Amano1983}, the bending mode $\nu_2$ \cite{Davies1984a,Kawaguchi1984} 
and C--O stretch $\nu_3$ \cite{Davies1984b}, quickly followed.
The radiative lifetime of \HCOp\ in the $\nu_1$ state \cite{Keim1990} and in the bending 
mode $\nu_2$ up to 8, has been studied using an ICR trap\cite{Mauclaire1995}.
Newer experimental studies, after 2000, focused on higher excited vibrational states, \eg, the determination
of $\nu_2$, $x_{22}$, $g_{22}$, and $B$(030) \cite{Foltynowicz2000}, and the acquisition of the pure rotational spectrum of \HCOp\ 
up to 1.2 THz \cite{Lattanzi2007}.
More recently, the $\nu_1$ band has been re-inspected in high resolution with the help of a 
frequency comb\cite{Siller2013}, and finally, \citet{Neese2013} reported hot band spectra of 
8 different vibrational modes of \HCOp.
Works focused on the higher energy isomer, \HOCp, are much more rare. The first rotational spectrum \cite{Gudeman1982}
has only been followed by the infrared detection of the $\nu_1$ band of \HOCp\ \cite{Nakanaga1987}, the 
hot band transition $\nu_1$ in $\nu_2=1$ \cite{Amano1990}, and another 
hot band study starting in the $\nu_2$ state by \citet{Amano2000}. 
Both \HCOp\ and \HOCp\ were subject to numerous theoretical spectroscopic studies \cite{Rogers1982,Botschwina1989,Martin1993,Mladenovic1998,Mladenovic2017,Kramer2010,Koput2019} with an emphasis 
on the prediction of the spectral properties of the lowest vibrational bands. While the calculated vibrational band 
origins are consistent across the studies, at least for the low lying states, the predicted lifetimes of 
the $\nu_1$ state of \HCOp\ differ by up to an order of magnitude.   

The abundance of \HOCp, \ie, the [\HCOp]/[\HOCp] ratio, has been observed in dense molecular clouds \cite{Apponi1997},
in diffuse clouds \cite{Liszt2004}, towards photodissociation regions (PDRs)\cite{Apponi1999,Fuente2003}, 
and towards the Galactic Center\cite{Abendano2015}.
It has also been used as a diagnostic tool, \eg, 
starburst energy feedback seen through \HCOp\ and \HOCp\ emission in NGC 253 \cite{Harada2021}.
Although the \HCOp/ \HOCp\ column density ratio has been found anywhere between 1000 and 9000 toward dark clouds, 
it has been observed to be significantly smaller (50-400) toward PDRs and diffuse clouds, suggesting an efficient \HOCp\ 
formation in these warmer environments.

In this work, we present a study of the first overtone vibrational transition of the C--H and O--H stretching 
mode of \HCOp\ and \HOCp. The measurements have been performed in a 
cryogenically cooled 22 pole ion trap using a laser induced reaction (LIR) action spectroscopy scheme. 
The setup allows us to determine the spectroscopic constants of the excited vibrational state for each ion as well 
as the corresponding radiative lifetimes and quenching rate coefficients with selected neutral partners.

\section*{Results and Discussion}
\label{results_discussion}

\subsection*{Overtone Spectroscopy of \HCOp/\HOCp}

The vibrational energy levels of both isomers are denoted by their vibrational quantum numbers, $\nu_1\nu_2^l\nu_3$, 
corresponding to the number of vibrational quanta in a given vibrational mode (following the notation by \citet{Neese2013}). 
As we probe the C--H/O--H stretching mode only, we further abbreviate 
this notation such that \vv\ has the same meaning as \Vvv.

The laser induced reaction (LIR) technique \cite{Schlemmer1999} relies on the different reactivities of the upper and 
the lower state involved in the transition. 
The \HCOp\ spectra were measured by using the endothermic reaction with \ce{CO2}  
\begin{equation}
    \HCOp\ + \ce{CO2} \xrightarrow[]{} \ce{HCO2+} + \ce{CO}, ~~~\Delta E = 0.55~\mathrm{eV},
\label{eq_HCOpLIR}
\end{equation}
and, similarly, using \ce{Ar} for \HOCp
\begin{equation}
    \HOCp\ + \ce{Ar} \xrightarrow[]{} \ce{ArH+} + \ce{CO}, ~~~\Delta E = 0.59~\mathrm{eV},
\label{eq_HOCpLIR}
\end{equation}
where the endothermicities were calculated based on the proton affinities of the reactants taken from 
NIST \cite{Lindstrom2023}. 
These processes are best recognized by following the reaction paths in the energy level diagram Fig.~\ref{fig_levels}. 
At room temperature and below, reactions (\ref{eq_HCOpLIR}) and (\ref{eq_HOCpLIR}) 
basically do not proceed due to the ions being almost exclusively in their ground vibrational state. 
The excitation of \HCOp\ or \HOCp\ to the \vv\ state and the ``activation'' of the reactions 
result then in an improved production of $\ce{HCO2+}$ or $\ce{ArH+}$ ions, respectively.    

\begin{figure}[h]
\centering
\includegraphics[]{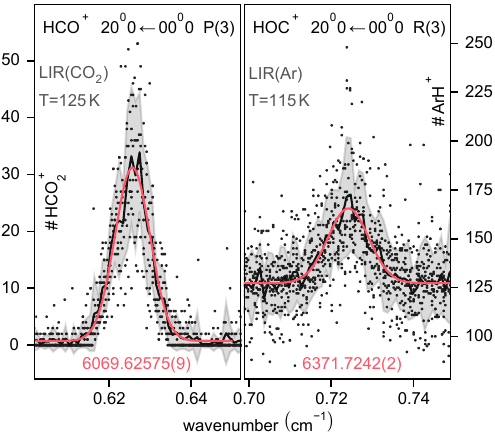}
\caption{P(3) and R(3) lines of \HCOp\ and \HOCp, respectively, measured by LIR technique in the 22 pole ion trap.
}
\label{fig_lines}
\end{figure}

Examples of obtained absorption lines are shown in Fig.~\ref{fig_lines}. The measured overtone transitions from 
the ground vibrational state to \vnu\ state for both \HCOp\ and \HOCp\ ions are summarized in 
Tab.~\ref{t_transitions}. 
The upper and lower levels can be described by a Hamiltonian \cite{Neese2013}
\begin{multline}
    H = T_{\nu} +B_\nu(J+1)-D_\nu\left[ J(J+1)\right]^2+\\
    H_\nu\left[ J(J+1)\right]^3,
    \label{eq_Hamiltonian}
\end{multline}
where $\nu$ stands for the vibrational quantum numbers of a given level, $T_\nu$ is the vibrational term energy, 
$J$ is the rotational quantum number, $B_\nu$ denotes the rotational constant, and $D_\nu$ and $H_\nu$ 
are the centrifugal distortion constants. 
The vibrational band origin for the \VNU\ transition is $\Wvv=T_{2\nu_1}-T_{0\nu_1}$. 

The measured transition wavenumbers were fitted using Hamiltonian (\ref{eq_Hamiltonian}), where 
the spectroscopic constants for the lower state were fixed at values experimentally determined by 
\citet{Neese2013} for \HCOp\ and by \citet{Amano2000} for \HOCp. 
The fitted spectroscopic constants are summarized in Tab.~\ref{t_sconstants} and compared to the previous 
theoretical and experimental studies in Tab.~\ref{t_sconstComp}. Note that two values, marked [N], acquired 
using cavity ring-down spectroscopy (CRDS), were excluded from the fit due to presence of other (unidentified) absorption features 
in immediate vicinity of the \HCOp\ transitions (see the dataset for more details).

The derived \HCOp\ spectroscopic constants are in a very good agreement with the hot band 
spectroscopy study by \citet{Neese2013} and also with recent quantum mechanical calculations \cite{Koput2019}. 
Although, in the \HOCp\ case, the older theoretical studies \cite{Martin1993,Mladenovic2017, Kramer2010} 
predicted the band origin \Wvv\ up to 30~\rcm\ away from the measured value,
the newer spectroscopic constants estimated by Koput\cite{Koput2019,Koput2021Priv_Comm}
are in good agreement with the experiment (the predicted band origin \Wvv\ is within two \Bvv\ from the experimental one).

\begin{table}[!h]
    \caption[]{Measured transition wavenumbers for the \VNU\ vibrational bands of \HCOp\ and \HOCp. 
    Statistical error reported in parenthesis. The absolute accuracy of the wavemeter is better than $0.002\;\rcm$.
    }\label{t_transitions}
    \begin{center}
    \begin{tabular}{lc|lc}
        \toprule
        \multicolumn{2}{c|}{\HCOp}        &  \multicolumn{2}{c}{\HOCp}   \\
        \midrule
        P(1)         &  6075.7093(1)            & R(1)  & 6366.0286(5)   \\
        \noalign{\smallskip}
        P(2)         &  6072.6892(1)            & R(2)  & 6368.8965(5)   \\
        \noalign{\smallskip}
        P(3)         &  6069.6258(1)            & R(3)  & 6371.7242(2)   \\
        \noalign{\smallskip}
        P(4)         &  6066.5171(1)            & R(4)  & 6374.5134(3)   \\
        \noalign{\smallskip}
        P(5)\TTT{c,N}&  6063.3620(9)            & R(8)  & 6385.2699(2)   \\
        \noalign{\smallskip}
        P(6)\TTT{c,N}&  6060.1689(2)            & R(9)  & 6387.8592(2)   \\
        \noalign{\smallskip}
        P(7)\TTT{c}  &  6056.9250(2)            & R(10) & 6390.4088(3)   \\
        \noalign{\smallskip}
        P(8)\TTT{c}  &  6053.6385(3)            & R(11) & 6392.9174(3)   \\
        \bottomrule
    \end{tabular}
    \vskip 0.2em
    \end{center}
    {\footnotesize{\textsf{All frequencies in \rcm. 
    Fitted temperature (\ie\ line width) varies between $130-210\;\tK$.
    [c] value obtained from the CRDS setup. [N] not included in the fit (see text). 
    }}}
\end{table}

\begin{table}[!h]
    \caption[]{Spectroscopic constants, radiative lifetimes \Tr, and quenching rate \kq{} for the $20^00$ vibrational state 
    of \HCOp\ and \HOCp\ ions derived in this work. 
    The vibrational transition moment for the \VNU\ transition and the reaction rate coefficients for the vibrational quenching of the \Vvv\ state in collision 
    with different gases were determined only for \HCOp\ ions.
    Statistical error is reported in parenthesis. }%
    \label{t_sconstants}
    \begin{center}
    \begin{tabular}{ll|l}
        \toprule
                       & \multicolumn{1}{c|}{\HCOp}  & \multicolumn{1}{c}{\HOCp}            \\
        \midrule
           $\Wvv$           & 6078.68411(19)          & 6360.17630(26)           \\
           $\Bvv$           & 1.465215(25)            & 1.472957(12)             \\
           $\Dvv$           & 3.56(44)$\cdot 10^{-6}$ & 4.03(11)$\cdot 10^{-6}$  \\ 
           $\Hvv$           & $-$                     & 3.3(2.2)$\cdot 10^{-10}$ \\
           $\langle 20^00|\mu|00^00\rangle$         
                            &  0.0084(25)\TTT{a}      &  $\mu_{\HCOp}/\,0.41(11)$\TTT{b} \\
           \Tr\ (ms)        &  1.53(34)               &  1.22(34) \\
           \kq{He}\TTT{c}   &  5.6(1.0)               &  $-$ \\
           \kq{H2}\TTT{c}   &  580(110)               &  $-$ \\
           \kq{N2}\TTT{c}   &  640(40)                &  $-$ \\
        \bottomrule
    \end{tabular}
    \vskip 0.2em
    \end{center}
    {\footnotesize{\textsf{Spectroscopic constants in \rcm.
    [a] lower estimate from the CRDS measurement in Debye. 
    [b] derived from relative ion signal intensities in the 22 pole trap, see text, 
    $R_{\mu}$, Eq.~(\ref{eq_A_ratio}).
    [c] in $10^{-12}\;\rateU$.
    }}}
\end{table}

\begin{table}[!h]
    \caption[]{Vibrational band origins and rotational constants of \HCOp\ and \HOCp\ obtained in previous 
    theoretical and experimental studies. 
    }\label{t_sconstComp}
    \begin{center}
    \begin{tabular}{lll}
        \toprule
        \multicolumn{1}{c}{$\Wvv$}  &  \multicolumn{1}{c}{$\Bvv$}    & Method/ Ref.  \\
        \midrule
        \multicolumn{3}{c}{\HCOp}\\
        6078.68411(19)             & 1.465215(25)   & E [present]   \\
        6078.6839\TTT{a}           & 1.4651993(13)  & D \cite{Neese2013}     \\
        6079.8                     & 1.46512        & T \cite{Koput2019}     \\
        \midrule
        \multicolumn{3}{c}{$\HOCp$} \\
        6360.17630(26)  & 1.472957(12)  & E [present]      \\
        6362.3          & 1.4736        & T \cite{Koput2021Priv_Comm}     \\
        6391.5\TTT{a}   & 1.4464\TTT{a} & T \cite{Martin1993}  \\
        6377.9          & 1.4707\TTT{a} & T \cite{Mladenovic1998,Mladenovic2017}  \\
        6390.5          &               & T \cite{Kramer2010}     \\
        \bottomrule 
    \end{tabular}
    \vskip 0.2em
    \end{center}
    {\footnotesize{\textsf{Spectroscopic constants in units of \rcm. 
    Method: T -- theory, E -- experiment, D -- derived from experimental data (fundamental + hot-band spectra).
    [a] Values calculated from published spectroscopic constants.
    }}}
\end{table}

\subsection*{Radiative Lifetime of the \vv\ State}

Measuring the overtone transitions while varying the number density of the LIR reactant allowed us to 
determine the radiative lifetime of the \vv\ state of \HCOp\ and \HOCp. 
In the steady state, with low laser power, the number of \HCOOp\ ions produced in the laser induced reaction (\ref{eq_HCOpLIR}) 
is proportional to the number of trapped \HCOp\ ions (predominantly in their ground vibrational state)
\begin{equation}
    N_{\HCOOp} = \frac{k_1[\COO]r_1 N_{\HCOp}}{k_1[\COO]+1/\Tr+\kq{M}[\ce{M}]}t,
\label{eq_quench}
\end{equation}
where $t$ is the irradiation time, $r_1$ corresponds to the rate of photon excitation to the \vv\ state,
$k_1$ to the reaction rate coefficient of vibrationally excited \HCOp\ with \COO, 
\Tr\ is the time constant of the radiative deexciation of the \vv\ state to lower lying vibrational levels,
and \kq{M} is the collisional reaction rate coefficient for the quenching of the \vv\ state in collision with particle $M$.
It is inherently assumed that $k_1$ is close to the Langevin collisional rate coefficient and every collision 
of the \vv\ state of \HCOp\ with \COO\ leads to \HCOOp, so the vibrational quenching induced by collisions with 
\COO\ can be neglected.

\begin{figure}[h]
\centering
  \includegraphics[width=80mm]{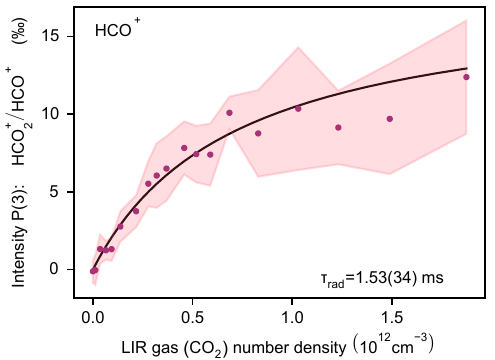}
  \caption{Determination of the $\tau_{rad}$ for \HCOp\ from the increase of the LIR signal as a function of \ce{CO2} 
  number density in the trap using Eq.~(\ref{eq_quench}).
  }\label{f_tauR}
\end{figure}

The radiative lifetime, \Tr, is then obtained by fitting equation (\ref{eq_quench}) to the dependence of the 
measured ratio of secondary (\HCOOp\ or \ArHp) to primary (\HCOp\ or \HOCp) ions as a function of the number 
density of the probing gas (\COO\ or \ce{Ar}) as shown in Fig.~\ref{f_tauR} for \HCOp\ ions. 
The resulting radiative lifetime was  $\Tr = 1.53 \pm 0.34$~ms for \HCOp(\vv) 
and $\Tr = 1.22 \pm 0.34$~ms for \HOCp(\vv) ions. 
The measurements were performed using the P(3) and R(3) lines for \HCOp\ and \HOCp, respectively, and the 
corresponding $k_1$ were calculated using polarisabilities from refs \cite{Asher1995,Alms2008}. Given that 
typical vibrational selection rules strongly favor transitions with vibrational quantum number changing by one, 
the measured radiative lifetimes probably pertain to the transition $\Vvv \rightarrow 10^00$
(note that the $10^00$ level is endothermic in reaction with \COO\ (for \HCOp) as well as \ce{Ar} (for \HOCp), \ie, it is not
probed in the selected LIR schemes).       

To our best knowledge, there are no experimental or theoretical data on the radiative lifetimes of the \vv\ states 
of either \HCOp\ or \HOCp. 
An approximate comparison with present data can be done applying the $1/\nu$ scaling law \cite{Mauclaire1995}, 
\ie, that the lifetime of the \vv\ state should be about half of the \vvone\ state. 
The predicted lifetimes for the \vvone\ state of \HCOp\ range from 0.4 ms \cite{Kramer2010} to 4.46 ms \cite{Rogers1982} 
with the majority of the studies closer to the upper value (e. g. 3.9 ms\cite{Botschwina1989} or 
3.7 ms \cite{Martin1993}). \citet{Keim1990} in their DLASFIB (direct laser absorption in a fast ion beam) 
experiment \cite{Keim1990} measured an absolute 
vibrational band intensity that corresponds to a lifetime of $5.9\pm0.9$~ms for the \vvone\ state of \HCOp. 
Considering the aforementioned $1/\nu$ scaling law, the value of \Tr\ for the \vv\ state of \HCOp\ measured in the
present experiment is thus within experimental error, comparable to that predicted for the \vvone\ state by 
theoretical calculations \cite{Rogers1982, Botschwina1989,Martin1993}, and lower than the value that could 
be extrapolated from the experiment by \citet{Keim1990}.

For \HOCp\ ions, only theoretical data on vibrational lifetimes are available. 
\citet{Kramer2010} predicted the lifetime of the \vvone\ state of \HOCp\ to be 0.4 ms, 
very close to the value they calculated for the same transition in \HCOp. Slightly higher values were reported 
by \citet{Rogers1982} (0.6 ms), \citet{Botschwina1989} (0.7 ms) and \citet{Martin1993} (0.7 ms). 
These results are substantially lower than 
the value obtained in present experiment with the lifetime of the \vv\ state of \HOCp\ re-scaled by the $1/\nu$ law. 
As this scaling law is valid for the harmonic approximation, it is difficult to distinguish whether the apparently 
longer than predicted radiative lifetime is due to anharmonic contributions to the wavefunctions or due to coupling to the 
other vibrational levels. Note that the scaling law is an approximation only, \eg, a deviation from the $1/\nu$ 
scaling by a factor of two was observed for the higher bending modes of \ce{DCO+} in an ion storage 
ring study \cite{Wester2002}.

\subsection*{Collisional Quenching of \HCOp(\vv)}

\begin{figure}[h]
\centering
  \includegraphics[width=80mm]{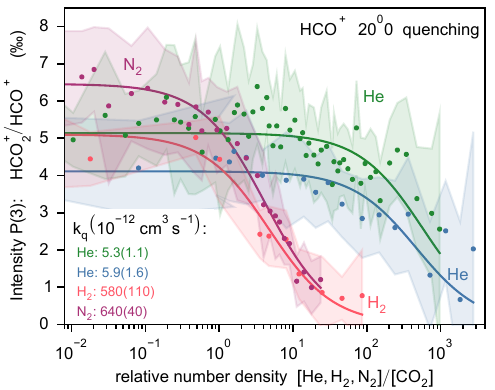}
  \caption{Determination of the quenching rate $k_q$ for \HCOp\ with \ce{He}, \ce{H2}, and \ce{N2} from the decrease of the LIR 
  signal as a function of the number density of the quenching partner in the trap using Eq.~(\ref{eq_quench}).
  The two different data sets for \ce{He}, acquired with $>20\,\%$ difference in the \ce{CO2} LIR reactant number density, 
  demonstrate the robustness of the technique.
  }\label{f_quench}
\end{figure}

In order to determine the rate coefficients for the quenching of the \vv\ vibrational state of \HCOp\ in collisions with different species, 
the number density of the probing gas (\COO) was kept constant and the number density of the collisional partner was varied over several 
orders of magnitude. 
The measured dependence of the ratio of secondary \HCOOp\ to primary \HCOp\ ions as a function of the quenching partner
number density at a temperature of 125~K is shown in Fig.~\ref{f_quench}. 
The quenching reaction rate coefficients \kq{\mathrm{M}} for the particular species M were obtained by fitting the data 
in Fig.~\ref{f_quench} using equation (\ref{eq_quench}) where \Tr\ was kept constant 
(at the value obtained in present study, $1.53 \pm 0.34$~ms). 
The resulting fitted reaction rates, summarized in Tab.~\ref{t_sconstants},
show that the quenching of the \vv\ vibrational state of \HCOp\ by molecular gases \ce{H2}, and \ce{N2}, is two orders of 
magnitude more efficient than quenching by atomic \ce{He}.
It is important to note that the number density of the LIR reactant (\ce{CO2}) has to be chosen so that the LIR
signal is high enough but at the same time not in saturation mode, \ie, somewhere in the first third of the curve
shown in Fig.~\ref{f_tauR}. In our quenching experiment, this value was ca. $3 \cdot 10^{11}\;\text{cm}^{-3}$. 
The sensitivity to the changes of \ce{CO2} number density (change by $20\,\%$) is shown on two different 
traces for \ce{He} quenching. The method seems to be reliable, as implied by Eq.~(\ref{eq_quench}), as long as the \ce{CO2}
number density is not too large.

When the quenching partner is a molecule, its rovibrational structure makes
vibrational energy transfer (V--V) possible \cite{Ferguson1986}, in addition to vibration-to-translation (V--T). 
This V--V transfer is a resonant process, and \HCOp\ possesses 
three normal modes with vibrational frequencies of $\nu_1 = 3089~\rcm$, $\nu_2 = 831~\rcm$, and $\nu_3 = 2183~\rcm$ \cite{Koput2019}, 
thus, there are many possible vibrational states where \HCOp\ can end up after a collision with a quenching gas 
(see Figure \ref{fig_levels}). Therefore, it can be expected that the measured quenching coefficients for collisions 
with molecular gases will have higher values than those that are typically obtained for vibrationally excited diatomic 
ions \cite{Bohringer1983, Ferguson1986}. In fact, the measuered quenching coefficients 
$\kq{H2} = 5.8\pm1.1 \cdot 10^{-10}~\rateU$ and $\kq{N2} = 6.4\pm0.4 \cdot 10^{-10}~\rateU$  are close to 
the corresponding Langevin collisional rates of $1.5\cdot 10^{-9}~\rateU$ and $8.2\cdot 10^{-10}~\rateU$, respectively.

In an ion trap study by \citet{Schlemmer2002}, the collisional quenching relaxation of 
antisymetric C--H stretch mode $\nu_3$ of \ce{C2H2+} with \ce{H2} has been estimated using a fit of a kinetic model 
to the number of reaction product as a function of \ce{H2} number density as $1.3\cdot10^{-9}\;\rateU$. Their approach
could be described as more sophisticated in comparison with the one presented in this work, as the 
applied LIR scheme \ce{C2H2+(\nu_{3}=1) + H2 \to C2H3+ + H} was not a simple proton transfer, 
\ie, the assumption of the Langevin behaviour did not hold 
and the reaction rate of the excited state had to be investigated concurrently. Their simulation also showed, that
the quenching rate with \ce{H2} was a ``substantial fraction ($0.81-0.87$) of the Langevin rate''.

Atomic helium seems to be surprisingly efficient at quenching the \vv\ state of \HCOp\ with 
$\kq{He} = 5.6 \pm 1.0 \cdot 10^{-12}~\rateU$. The weak helium-ion interaction potential often results 
in small ($<10^{-15}~\rateU$ at 300~K) deexcitation rates as was observed, \eg, for 
\ce{N2+} and \ce{NO+} \cite{Ferguson1986,Kato1995}. Even smaller reaction rate coefficients were predicted for low 
temperature collisional quenching of diatomic anions (\ce{CN-}, \ce{C2-}) by helium \cite{Mant2020,Mant2021}. 
\citet{Wisthaler2000} reported a rate coefficient of 
$1.3 \cdot 10^{-13}~\rateU$ for the quenching of vibrationally excited \ce{HCN+} and \ce{DCN+} ions in collision 
with helium at a collisional energy corresponding to a temperature of 1500~K, noting that the quenching reaction 
rate coefficient fell down below $10^{-14}~\rateU$ at lower energies.

The \HCOp--He potential energy surface has a minimum of approximately 300 \rcm\ below the dissociation limit \cite{Tonolo2021}. 
For comparison, the \ce{N2+}--He system has minimum of only 140 \rcm\ \cite{Miller1988} (with a corresponding vibrational quenching 
reaction rate coefficient of $1\cdot10^{-15}~\rateU$) and the \ce{O2+}+Ar system has a lowest dissociation energy of 
765 \rcm\ \cite{Catani2023} and the reported quenching coefficient for the $\nu = 1$ state of \ce{O2+} was 
$1\cdot 10^{-12}~\rateU$ \cite{Bohringer1983}. We note that it is very hard to predict the magnitude of the quenching 
coefficient based only on the interaction potentials and the polarisability of the neutral without a full quantum 
mechanical calculation \cite{Mant2021}.

\subsection*{Transition Dipole Moments for \VNU\ Transition of \HCOp\ and \HOCp}

In order to estimate the value of the integrated molar infrared intensity of 
absorption $\gamma$ for the \VNU\ transition of \HCOp\ in 
the Stationary Afterglow with Cavity Ring-Down Spectrometer (SA-CRDS) 
apparatus, we have followed the approach used for \ce{H3+} ions by \citet{Shapko2021}. 
Aided by a chemical kinetics model,
the experimental conditions were set to maximize the amount of \HCOp. 
The measured absorption signal can then be compared to the value of the electron number density determined at the same 
time in the afterglow by microwave diagnostics. Using the P(4), P(7) and P(8) transitions of \HCOp, 
the obtained lower estimate for the molar infrared intensity of 
absorption was $\gamma = 18\pm5~\cmsmol$, corresponding to a vibrational transition moment $\langle 20^00|\mu|00^00\rangle = 0.0084\pm0.0025$ D. 
This is lower than the value of $\gamma = 23.7~\cmsmol$ predicted by \citet{Botschwina1989}. 
The measurements were performed at temperatures of 80 K and 170 K and the \HH\ and \ce{CO} number densities were varied in 
the range of $5\cdot10^{13} - 5\cdot10^{15}\;\text{cm}^{-3}$ with helium buffer gas pressure in the range of 500 to 1000 Pa. 
The stated error arises mainly from the estimated uncertainty of the electron number density determination.     

\medskip
Although it is inherently difficult to derive absolute intensities in an action spectroscopy experiment in an ion trap,
relative evaluation is possible. We show how to correctly assess the relative intensities of
two different species in a LIR action spectra in Eqs.~(\ref{eq_A_derivation}--\ref{eq_Arot_to_murot}). 
For high number densities of LIR probing gas and no quenching, equation (\ref{eq_quench}) simplifies to
\begin{align}
    N_{\HCOOp}/N_{\HCOp} &= \rho_1 = r_1\left(\HCOp\right) t,~~~~~\text{for}~ \HCOp \label{eq_quench_sat} \\
    N_{\ArHp}/N_{\HOCp}  &= \rho_2 = r_1\left(\HOCp\right) t.~~~~~~\text{for}~ \HOCp \nonumber
\end{align}
The measured values were $\rho_1 = 0.0099\pm0.0013$ from the P(3) line of \HCOp\ at 125~K, and $\rho_2 = 0.025\pm0.005$ 
from the R(3) line of \HOCp\ at 115~K, at the same trapping time $t=1.7\;\text{s}$. 
In the latter case, the chemical probing by \iNN\ has shown that only ca. 70 \% 
of all ions with mass 29 \massU\ were \HOCp, \ie, $\rho_2$ has to be increased accordingly.
Assuming that Herman-Wallis factors are close to unity for both \HCOp\ and \HOCp, the resulting calculated 
ratios for the Einstein A coefficients (Eq.~(\ref{eq_A_ratio})) and vibrational transition moment squared (Eq.~(\ref{eq_Arot_to_murot})) 
are $R_A = 0.48\pm0.10$ and $R_\mu = 0.41\pm0.11$. 
The stated uncertainty is only the statistical error. Systematic uncertainties arise mainly from the completely unknown overlap of 
the laser beam (different light sources/optics/path for both species) with the ion cloud in the trap.
Even thought \HCOp/\HOCp\ have the same \massU, the distribution of the ions inside the trap will most probably differ,
since neither effective potential $V^*$, nor the number of ions were the same, and on top, in case of \HOCp, also the presence 
of the dominant lighter \ce{C+} ion will play a significant role. Therefore, the results of this relative intensity 
determination technique are only illustrative, as we are unable to provide a reliable confidence interval.

\section*{Conclusions}
\label{conclusion}

We present the first spectra of the \vv\ overtone transition of \HCOp\ and \HOCp. The availability of this data
and cheap telecommunication laser diodes opens up the possibility of very cheap monitoring of these two isomers 
in any optically thin environment or in emission.
Although the accuracy of the theoretical prediction of the \HOCp\ band origin by \citet{Koput2021Priv_Comm} is in the \rcm\ range, 
we hope that the experimental spectroscopic constants will be helpful for further improvement of the \textit{ab initio}
theoretical models.

We explore the versatility of the cryogenic ion trapping technology and present a technique to improve 
the LIR action spectroscopy for very weak signals into a background free action spectroscopic method,
by actively removing the background in the ion signal by selective \massU\ ejection from the trap. 
Furthermore, we extensively describe the method of determination of radiative lifetimes, \Tr, and quenching reaction rates, $k_q$,
and its caveats inside an ordinary 22 pole rf trap, easily applicable on a plethora of molecular ions.

Although similar experiments have been done in the past, \eg, \Tr\ in an ICR \cite{Mauclaire1995, Heninger2003}, 
or measurements of quenching reaction rates for various, mainly diatomic, ions \cite{Bohringer1983,Kato1995,Kato1998,Mant2021}, 
this data exist only for a select few systems. Moreover, there are not many experiments focused on higher lying vibrational levels.
The collisional quenching rates are needed for the treatment of the molecular emission lines in astronomy. 
We aspire to stimulate further study of \Tr, and quenching rates for more ion molecular systems, as well as for the
different transitions, \eg, different vibrational modes (bending, stretching, fundamentals, overtones etc.).

Last but not least, our quenching technique only shows the removal of the vibrational excitation in the molecular ion. 
We can not estimate what happened with the energy, whether it redistributed towards translational energy or into internal 
excitation of the quenching neutral molecule. These fundamental ion-molecule processes are not only of interest for
astrophysics, but also closely related to action spectroscopy technology, \ie, LOS (leak-out spectroscopy) 
scheme\cite{Schmid2022}, where only the kinetic energy release is monitored in a quenching event. Consequently,
a study of one system by both techniques could provide a full picture of energy distribution in the de-exciting collision.

\section*{Experimental Section}
\label{experimental}

The measurements have been predominantly conducted in the CCIT 22 pole rf cryogenic trap setup \cite{Jusko2023}, 
with the exception of the P(5)--P(8) transitions of the \HCOp\ ion acquired in the CRDS setup\cite{Hlavenka2006}. 
The experimental setups are extensively described in the references provided, we will only focus on particular
improvements of the 22 pole trap setup needed for this experiment (the CRDS setup has been used without any modifications). 

\subsubsection*{22 Pole Trap Setup}

The 22 pole trap setup CCIT\cite{Jusko2023} consists of a storage ion source, quadrupole mass filter used to select the mass
of the ion to be stored in the trap, the 22 pole trap mounted on variable temperature cryostat and a mass sensitive detection system. 
Each data point in the experiment, \ie, ion production, trapping, storage/ irradiation/ LIR reaction and detection, 
is acquired in a 3 s cycle time.

\begin{figure}[tb]
\centering
\includegraphics[width=80mm]{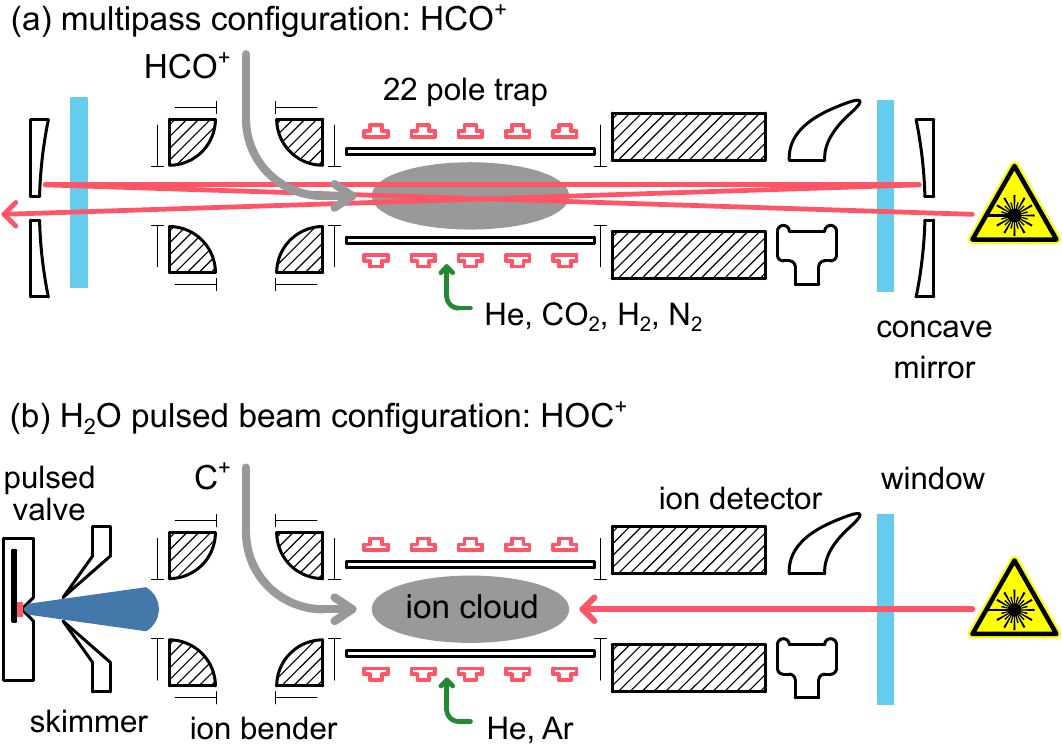}
\caption{Experimental setup in a laser multi-pass configuration (\HCOp\ experiments -- (a)), 
and an on axis molecular beam configuration (\HOCp\ experiments -- (b)).
Not to scale. For a detailed description of the experiment see ref. \cite{Jusko2023}}
\label{fig_exp}
\end{figure}

\HCOp\ was produced directly  from \ce{CO} and \ce{H2} precursor gases using electron bombardment in the 
storage ion source\cite{Gerlich1992}, where the produced ions have many collisions with both \ce{CO} and \ce{H2}, 
assuring dominant production of the lowest isomer. 
Subsequently, \HCOp\ was mass selected in a quadrupole filter and injected into the trap, 
where a short intense \ce{He} pulse was used to trap it and provide collisional
cooling into the ground state (see Fig.~\ref{fig_exp}(a)).

In order to produce \HOCp in sufficient quantities, the following ion-molecule reaction\cite{Sonnenfroh1985} inside the trap was used
\begin{align}
    \ce{C+} + \ce{H2O} &\xrightarrow[]{} \ce{HOC+} + \ce{H}, ~~~~\Delta E = -2.60~\mathrm{eV},\label{eq_CpHHO}\\
                       &\xrightarrow[]{} \ce{HCO+} + \ce{H}, ~~~~\Delta E = -4.34~\mathrm{eV}.
\end{align}%
The \ce{C+} ion was produced in the ion source from \ce{CH4} precursor gas using electron bombardment, mass selected
and injected into the trap, where a short intense \ce{He} pulse was used to trap and provide collisional cooling. 
Subsequently, an intense water pulse from the side of the trap (on the axis, ca. 200~ms long) was
used to convert approx. 1/3 of the \ce{C+} ions into mass 29~\massU\ (\HCOp\ and \HOCp) (see Fig.~\ref{fig_exp}(b)).
The on axis water beam had to be used to avoid the freezing of the water molecules at temperatures $< 200$~K.
The ratio of the \HOCp\ to \HCOp\ ions was determined using proton transfer to \iNN\ as a chemical probing reaction. 
This reaction is exothermic for the \HOCp\ reactant only (see Fig.~\ref{fig_levels}. 
Note that \ce{^{14}N2H+} has the same mass as \HOCp).
This method of \HOCp\ production and isomer fraction determination has been recently described by \citet{Yang2021}.
The gas pulses into the trap, as well as on the axis, are delivered using a custom built piezo element actuated valves, 
operated at the resonant frequency of a few kHz.

In both cases, following the ion preparation/ trapping, the ions are left stored in the trap for a predetermined time 
(1.75 s), during which the IR radiation can excite the ions in the ground state. The probing reactant gas (\ce{CO2} in the case of \HCOp\
or \ce{Ar} in the case of \HOCp\ spectroscopy) is flowing directly into the trap freely, maintaining a constant number density 
therein during the whole cycle time. 
Neither of the products of the LIR processes (\ref{eq_HCOpLIR}) and (\ref{eq_HOCpLIR}), \ce{HCO2+} or \ce{ArH+}, can
undergo any significant further reaction inside the trap and thus accumulate until being detected at the end of the cycle.
In the spectroscopy experiment, the wavelength of the laser radiation is scanned by a small step after every single data point taken and the
spectrum as seen in Fig.~\ref{fig_lines} is acquired.

In case of the radiative lifetime experiment and quenching experiment (Figs.~\ref{f_tauR}, \ref{f_quench}), the light wavelength
is kept constant (on resonance), while the number density of the probing gases has to be varied using a variable leak valve.
The pressure inside the vacuum vessel has been recorded using a Bayard-Alpert ion gauge (calibrated by an absolute pressure 
baratron gauge), the temperature transpiration effect was taken into account to accurately determine the number 
density in the trap (for details see ref.\cite{Jusko2023}).

We exclusively used continuous wave fiber pigtailed telecommunication grade distributed feedback (DFB) laser diodes 
FLPD-1647-07-DFB-BTF (\HCOp\ P(1)--P(4)), QDFBLD-1650-5 (\HCOp\ P(5)--P(6)),
NX8570SD654Q-55 (\HOCp\ R(8)--R(11)), and QDFBLD-1570-20 (\HOCp\ R(1)--R(4)). A small part of the light has been split  
(fiber beam splitter, $1\,\%$) and fed into the WA-1650 (EXFO) wavemeter, with an absolute accuracy better than 0.3~ppm at 1500~nm.
The majority of the light was guided into the trap using an adjustable focus collimator. 
Since the laser diode power for the \HCOp\ experiment was only ca. 7~mW, we decided to use a Pfund cell 
(two spherical mirrors $f=750$~mm with a central $\phi 1.5$~mm hole) in a 3-pass configuration,
effectively multiplying the power inside the trap by a factor of $\approx 2.5$ (see Fig.~\ref{fig_exp}(a)). 
The radiation power has been regularly monitored with a Ge photodiode.

\subsubsection*{Mass Selective Ejection -- Zero Background Spectroscopy}

The ions coming from the source can easily be mass selected to contain only the \massU\ ion of interest.
The ion (\HCOp) is consequently injected into the trap, where the stopping \ce{He} pulse slows the ion down and removes its internal
energy. However, if some other gas, \eg, probing gas (\ce{CO2}), is present in the trap, some of the incoming ions
will collide with it prior to thermalisation (kinetic, internal degrees of freedom) and undergo a chemical 
reaction (product \HCOOp). 
As a consequence, it is impossible to fill the trap containing a probing gas and not produce any of the ions which are detected 
in the LIR scheme. This is usually of no concern for fundamental vibrational spectroscopy, where the absorption rates are usually 
two orders of magnitude stronger than for overtone vibrational transitions and the ``injection'' only creates 
a few percent of the total ions detected in the LIR scheme.

Fortunately, in this work, the products of the LIR reaction (\ce{HCO2+}: $45\;\massU$, \ArHp: $41\;\massU$) 
are significantly heavier than the ions of interest $29\;\massU$. 
Therefore, we can vary the effective potential $V^*$ in the trap, by lowering
the driving amplitude $V_0$ for few hundred of ms, to achieve conditions where the higher mass ion is no longer efficiently trapped,
effectively cleaning the trap from masses higher than a set threshold of ca. 40 \massU\ 
(see \citet{Gerlich1992} for details on trap technology).
The effect of this technique can be best observed on the left panel of Fig.~\ref{fig_lines}, where the off resonance background
signal of \HCOOp\ is essentially 0 (note that the value is not kept at 0 on purpose, since it is difficult to spot any instrument 
malfunction while scanning without having any signal at all).

The same technique is applied for the \HOCp\ spectroscopy, where the few hundred ms long ejection time additionally allows the 
excited \HOCp/\HCOp,
produced in a strongly exothermic reaction~(\ref{eq_CpHHO}), to radiatively decay to its vibrational ground state.
Unfortunately, in this configuration, some of the water vapor and precursor ion \ce{C+} is always present, effectively
creating excited products of the reaction~(\ref{eq_CpHHO}) all the time. These can further react with the probing neutral \ce{Ar},
even off \HOCp\ resonance, and contaminate the trap with \ce{ArH+} during the 1.75~s irradiation time.
To provide some quantitative values, in case of \HOCp\ experiment, the trap is operated at $f_0=19.5\,\text{MHz}$
and $V_0$ from $41$~V (everything trapped) to $26$~V (ejection of masses $>40\;\massU$). The change in effective potential 
$V^*(0.8\,r_0)$ for mass $29\;\massU$ is then $5.2 \to 2.1\,\text{meV}$ 
and respectively for mass $41\;\massU$ it is $3.7 \to 1.5\,\text{meV}$.

\subsubsection*{CRDS Setup}

The SA-CRDS apparatus (Stationary Afterglow with Cavity Ring-Down Spectrometer) was utilized for the 
measurement of 4 line positions in this study. Originally developed for the  determination of recombination rate coefficents 
for collisions of electrons with molecular ions \cite{Macko2004,Shapko2020}, it was also used for measurements of the line 
positions of overtone transitions for several polyatomic ions \cite{Hlavenka2006,Kalosi2017}.

The main diagnostic technique is a continuous wave modification of cavity ring-down spectroscopy based on a design 
by \citet{Romanini1997}. The \HCOp\ ions were produced in a pulsed microwave discharge (discharge power 
of $\approx$ 17 W, period of 4.9~ms, 40\% duty cycle) ignited in a He/Ar/\HH/CO mixture with reactant number 
densities of $8\cdot10^{17}/2\cdot10^{14}/4\cdot10^{14}/4\cdot10^{14}$ $\text{cm}^{-3}$. The discharge tube made of fused 
silica was immersed in liquid nitrogen. 

We employed a model of chemical kinetics and varied the reactant number densities  by almost an order of magnitude to maximize the number of \HCOp\ ions in the discharge and afterglow plasma.

\subsection*{Radiative Deexcitation/ Collisional Quenching Derivation}
\label{sec_radiative_quenching}

In the following, we derive Eq.~(\ref{eq_quench}), used for the determination of \Tr\ and $k_q$, for the \HOCp(\vv) case. The technique can be applied on any analogous LIR system and transition (fundamental, overtone etc.), 
\eg, for \HOCp\ ions with Ar probing gas, etc.
For simplicity, let's consider only the most important processes involving \HCOp\ ions in the 22 pole rf trap:

1. Overtone excitation
\begin{equation}
    \HCOp(\vvnull) + h\nu \xrightarrow[]{r_1} \HCOp(\vv),
\end{equation}
where $r_1$ is the rate for excitation from the vibrational ground state by two vibrational quanta in the 
stretching mode $\nu_1$. It depends on the corresponding transition line strength, the intensity of the laser and on the
overlap between the ion cloud in the trap and the laser beam.

2. Deexcitation of the vibrationally excited \ce{HCO+} ions to lower states 
\begin{equation}
    \HCOp(\vv) \xrightarrow[]{1/\tau} \HCOp(\vvnull,\vvone) + h\nu,
\end{equation}
here $1/\tau = 1/\Tr + 1/\Tq$, with \Tr\ representing spontaneous emission, and \Tq\ deexcitation due to collisions 
with another particle. 
In our particular case, it can be safely assumed that \Tr\ is mainly given by the transition 
to the \vvone\ state.

3. Proton transfer in reaction with the probing gas
\begin{equation}
    \HCOp(\vv) + \COO \xrightarrow[]{k_1} \HCOOp + \ce{CO}.
\end{equation}
The corresponding rate equations for \HCOp(\vv) and \HCOOp\ can then be written as
\begin{align}
    \frac{\mathrm{d}N_{\HCOp(\vv)}}{\mathrm{d}t} &= -k_1N_{\HCOp(\vv)}[\COO] - \frac{1}{\tau} N_{\HCOp(\vv)} +\nonumber \\ 
                                                 &  ~~~~r_1N_{\HCOp(\vvnull)},   \label{eq_HCO+v2} \\
    \frac{\mathrm{d}N_{\HCOOp}}{\mathrm{d}t}     &= k_1N_{\HCOp(\vv)}[\COO], \label{eq_HCO2+}
\end{align}
where $N_\mathrm{i}$ denotes number of ions of species $i$ in the trap.
\medskip %

By utilising the steady-state approximation, we can express $N_{\HCOp(\vv)}$ from equation (\ref{eq_HCO+v2}) and 
insert it into equation (\ref{eq_HCO2+}). 
After simple integration we obtain 
\begin{equation}
    N_{\HCOOp} = \frac{k_1[\COO]r_1N_{\HCOp(\vvnull)}}{k_1[\COO]+1/\tau}t
\label{eq_tau_rad}
\end{equation}
for the number of detected \HCOOp\ ions, where $t$ is trapping time.
Noting that $1/\tau = 1/\Tr + 1/\Tq$ and that $1/\Tq = \kq{\mathrm{M}}[\mathrm{M}]$, equation (\ref{eq_tau_rad}) can be rewritten as
\begin{equation}
    N_{\HCOOp} = \frac{k_1[\COO]r_1N_{\HCOp(\vv)}}{k_1[\COO]+1/\Tr+\kq{\mathrm{M}}[\mathrm{M}]}t,
\label{eq_quench2}
\end{equation}
where \kq{\mathrm{M}} is the reaction rate coefficient for the quenching of the \vv\ state of \HCOp\ in collisions with 
particles $\mathrm{M}$. 
This relation can be used to determine the lifetime of spontaneous emission by varying the LIR reactant number density,
as well as to evaluate the quenching rate by varying the number density of the quenching particles $[\mathrm{M}]$ 
(provided \Tr\ is known), from the measured LIR signal, $N_{\HCOOp}/N_{\HCOp(\vvnull)}$, as a function of the respective number density.

\subsection*{Relative Transition Intensities in Action Spectroscopy} %
\label{sec_rel_inten}

In saturation conditions, \ie,
conditions with high LIR reactant gas number density such that neither \Tr\ nor quenching play a role, the intensity of the LIR signal, defined by Eq.~(\ref{eq_quench2}), reduces to Eq.~(\ref{eq_quench_sat}), 
\ie, $\rho = r_1 t$.
The rate of excitation to the \vv\ vibrational state is proportional to the laser power and to the Einstein coefficient for given transition. 
Following the derivation in refs. \cite{Rothman1998,Shapko2021} and assuming a thermal population of states, the Einstein coefficient 
$A_{JvJ'v'}$ for spontaneous emission between two rovibrational states of \HCOp\ is  proportional to the measured ratio $\rho$ of the numbers 
of secondary (\HCOOp\ or \ArHp) and primary (\HCOp\ or \HOCp) ions obtained at high probing gas number densities
\begin{equation}
    A_{JvJ'v'} \sim \rho \frac{8\pi\nu^2cQ(T)}{g_\mathrm{u}} \exp{\frac{E}{k_\mathrm{B}T}}\sqrt{\frac{\pi}{4\ln{2}}}w/I,
\label{eq_A_derivation}
\end{equation}
where $\nu$ is the transition wavenumber, $c$ is speed of light, $k_\mathrm{B}$ is the Boltzmann constant, $Q(T)$ is the partition function at 
temperature $T$, $g_\mathrm{u}$ denotes the statistical weight of the upper state, $E$ is the energy of the lower state of the transition, 
$w$ is the full width at half maximum (FWHM) of the Doppler broadened line, 
and $I$ is the number of photons passing through the trap per second. 
The ratio between the Einstein coefficients of two rovibrational transitions is then
\begin{equation}
    R_A=\frac{A_1}{A_2}=\frac{\rho_1\nu_1^2w_1g_\mathrm{l1}g_\mathrm{u2}P_2I_2}{\rho_2\nu_2^2w_2g_\mathrm{l2}g_\mathrm{u1}P_1I_1},
\label{eq_A_ratio}
\end{equation}
where the subscripts 1 and 2 distinguish between the transitions and $P$ is the thermal population of the lower state with statistical 
weight $g_\mathrm{l}$. The state to state transition dipole moment squared $\langle Jv|\mu|J'v'\rangle^2$ is connected to the corresponding 
vibrational transition moment $\langle v|\mu|v'\rangle$ through the relation \cite{Bernath2005}
\begin{equation}
    \langle Jv|\mu|J'v'\rangle^2=\langle v|\mu|v'\rangle^2S^{\Delta J}_JF(m),
\label{eq_muvib_to_murot}
\end{equation}
where $S^{\Delta J}_J$ and $F(m)$ are H{\" o}nl-London and Herman-Wallis factor, respectively. Given that \cite{Rothman1998}
\begin{equation}
    A_{JvJ'v'} = \frac{64\pi^4}{3h}\nu^3 \frac{g_\mathrm{l}}{g_\mathrm{u}}\langle Jv|\mu|J'v'\rangle^2\times10^{-36} ~\mathrm{s^{-1}},
\label{eq_Arot_def}
\end{equation}
with $h$ in cm$^2$gs, $\nu$ in \rcm\ and $\langle Jv|\mu|J'v'\rangle$ in Debye, it follows that
\begin{equation}
    R_\mu=\frac{\langle v|\mu|v'\rangle^2_\mathrm{HCO^+}}{\langle v|\mu|v'\rangle^2_\mathrm{HOC^+}} = 
    \frac{\rho_1w_1\nu_2P_2I_2S^{\Delta J}_{J2}F_2(m)}{\rho_2w_2\nu_1P_1I_1S^{\Delta J}_{J1}F_1(m)}.
\label{eq_Arot_to_murot}
\end{equation}

\section*{Acknowledgements}
This work was supported by the Max Planck Society, 
the Czech Science Foundation (Grant Nos. GACR 23-05439S and GACR 22-05935S), 
and the Charles University (Grant Nos. GAUK 337821 and GAUK 332422). L.U. acknowledges 
support by the COST Action CA21101.
The authors gratefully acknowledge the work of the electrical and mechanical workshops and 
engineering departments of the Max Planck Institute for Extraterrestrial Physics.

\section*{Conflict of Interest}
The authors report there are no competing interests to declare.

\section*{Additional Information}
The data that support the findings of this study are openly available in Zenodo at 
\url{https://doi.org/10.5281/zenodo.10473481}, reference number 10473481.
\begin{shaded}
\noindent\textsf{\textbf{Keywords:} \keywords} 
\end{shaded}
\bibliographystyle{rsc}
\bibliography{lit}

\end{document}